\documentstyle[fancybox,epsfig,color,axodraw,12pt]{article}

\def\plb#1{Phys.~Lett.~{\bf B#1}}
\def\npb#1{Nucl.~Phys.~{\bf B#1}}
\def\prl#1{Phys.~Rev.~Lett.~{\bf #1}}
\def\prd#1{Phys.~Rev.~{\bf D#1}}

\let \nn = \nonumber

\def\mt{{\ifmmode M^{eff}_T\else $M^{eff}_T$\fi}}

\def\e{\epsilon}

\def\l{\left}
\def\r{\right}

\def\ln#1{\mbox{ln}\l(#1\r)}

\def\e3{$\epsilon_3$}

\def\ch2{$\chi^2$}

\def\co#1{{\ifmmode{\cal O}_{#1}\else${\cal O}_{#1}$\fi}}

\newdimen\unit
\def\point#1 #2 #3{\vbox to0pt{\kern-#2\unit
  \hbox{\kern#1\unit#3}\vss}
 \nointerlineskip}

\newcommand{\be}{\begin{equation}}
\newcommand{\ee}{\end{equation}}
\newcommand{\bea}{\begin{eqnarray}}
\newcommand{\eea}{\end{eqnarray}}

\begin{document}
\thispagestyle{empty} \noindent
\begin{flushright}
        October 2002
\end{flushright}

\vspace{1cm}
\begin{center}
Proton Decay\footnote{Talk presented at SUSY'02, DESY, Hamburg,
Germany}
\end{center}
  \vspace{1cm}
    \begin{center}
S. Raby\\
      \vspace{0.3cm}
\begin{it}
Department of Physics, The Ohio State University, \\ 174 W. 18th
Ave., Columbus, Ohio  43210
\end{it}
  \end{center}
  \vspace{1cm}
\centerline{\bf Abstract}
\begin{quotation}
\noindent  We discuss the status of supersymmetric grand unified
theories [SUSY GUTs] with regards to the observation of proton
decay.  In this talk we focus on SUSY GUTs in 4 dimensions. We
outline the major theoretical uncertainties present in the
calculation of the proton lifetime and then present our best
estimate of an absolute upper bound on the predicted proton
lifetime.  Towards the end, we consider some new results in higher
dimensional GUTs and the ramifications for proton decay.
\end{quotation}
\vfill\eject

\section{Introduction}

Preliminary Super-K bounds \cite{kearns} on the proton and neutron
lifetimes provide stringent constraints on grand unified theories
\cite{guts,so10}.
$$\begin{array}{|c|c|c|}
\hline
{\rm mode} &  {\rm exposure} & \tau/{\rm B \;\; limit} \\
            &     ({\rm kt} \cdot {\rm yr}) &  (10^{32} \;{\rm yrs}) \\
\hline
p  \rightarrow  \pi^0 + e^+   &  79 & 50 \\
p  \rightarrow  \pi^0 + \mu^+   &  79 & 37 \\
p  \rightarrow  K^+ + \bar \nu   &  79 & 16 \\
p  \rightarrow  K^0 + e^+   &  70 & 5.4 \\
p  \rightarrow  K^0 + \mu^+   &  70 & 10 \\
\hline
n   \rightarrow  K^0 + \bar \nu   &  79 & 3.0 \\
\hline
\end{array}$$

These constraints place bounds on the GUT scale.  For example, a
generic 4 Fermion baryon and lepton number violating operator of
the form $\frac{1}{\Lambda^2} \;\; q \ q \ q \ l$ results in a
proton decay rate, typically of order $\Gamma_p \sim 10^{-3} \;
m_p^5/\Lambda^4$. Thus a bound on the proton lifetime $\tau_p
> 5 \times 10^{33} {\rm yrs}$ roughly constrains the scale
$\Lambda > 4 \times 10^{15}$ GeV.

4D SUSY GUTs have many notable virtues.
\begin{itemize}
\item GUTs \cite{guts,so10} explain the standard model charge
assignments for the observed families. \item  They predict the
unification of the three gauge couplings at the GUT scale
\cite{eft} and the prediction of SUSY GUTs
\cite{susygut,sguttwo,gutexp} agrees quite well with the low
energy data. \item Bottom-Tau or Top-Bottom-Tau Yukawa coupling
unification is predicted in simple $SU_5$ or $SO_{10}$, resp.
\item Including additional family symmetries relating different
generations leads to simple models of fermion masses and mixing
angles. \item Neutrino oscillations governed by a see-saw scale of
order $10^{-3} \ M_G$ are easily included.  \item  The lightest
SUSY particle [LSP] is a natural dark matter candidate, and \item
SUSY GUTs provide a natural framework for understanding
baryogenesis and/or leptogenesis.
\end{itemize}

In order to set the notation recall, in $SU(5)$, quarks, leptons
and Higgs fields are contained in the following GUT
representations\cite{guts}: $\{ Q = \left(\begin{array}{c} u
\\ d \end{array}\right) \;\;\;  e^c \;\;\; u^c \} \;\; \subset
\;\; \bf 10$, $ \{  d^c \;\;\; L = \left(\begin{array}{c} \nu \\ e
\end{array}\right) \} \;\; \subset \;\; \bf \bar{5}$ and $ \left( H_u, \ T
\right),\;\; \left( H_d, \ \bar T \right) \;\; \subset \;\; {\bf
5_H},\;\; {\bf \bar{5}_H} $.  While for $SO_{10}$\cite{so10} we
have all quarks and leptons of one family (including one
additional state, a ``sterile" neutrino) ${\bf 10} + {\bf \bar{5}}
+ \bar{\nu}_{sterile} \;\; \subset \;\; \bf 16 $ and the Higgs in
one irreducible representation $ {\bf 5_H},\;\; {\bf \bar{5}_H}
\;\; \subset \;\; {\bf 10_H} $.

At the moment, the only experimental evidence we have for low
energy SUSY comes from gauge coupling unification
\cite{susygut,sguttwo,gutexp}.
\begin{center}
\SetPFont
{Helvetica}{15}
\begin{picture}(300,45)(0,0)
\SetColor{Red} \Line(180,28)(50,15) \Line(180,25)(50,35)
\Line(180,25)(50,45) \Text(35,10)[1c]{$\alpha_3^{-1}$}
\Text(35,30)[1c]{$\alpha_2^{-1}$}
\Text(35,45)[1c]{$\alpha_1^{-1}$}

\SetColor{Black} \Line(180,26)(220,22)
\Text(227,30)[1c]{$\alpha_G^{-1}$} \Text(50,0)[1c]{$M_Z$}
\Text(180,0)[1c]{$M_G$}

\SetColor{Black}
\end{picture}
\end{center}
The current status of this analysis uses two loop renormalization
group running from $M_G$ to $M_Z$ with one loop threshold
corrections at the weak scale.  It is important to note that there
are significant one loop, GUT scale, threshold corrections from
the Higgs and GUT breaking sectors.   Hence at two loops the three
gauge couplings do not meet at $M_G$. Nevertheless, the GUT scale
can be defined as the point where two couplings meet;
$\alpha_1(M_G) = \alpha_2(M_G) \equiv \tilde \alpha_G$.  Then we
define $\epsilon_3 \equiv \frac{(\alpha_3(M_G) - \tilde
\alpha_G)}{\tilde \alpha_G}$.   A negative 4\% correction at the
GUT scale is sufficient to precisely fit the low energy data.

\section{Nucleon Decay}

In SUSY GUTs, nucleon decay is affected by dimension 4, 5 and 6
operators.

$\bullet$ For dimension 6 operators $\Lambda \approx M_G$ and for
$M_G \sim 3 \times 10^{16} $ GeV we find $\tau_p \sim 10^{35 \pm
1}$ yrs. with the dominant decay mode $p \rightarrow \pi^0 + e^+$.

$\bullet$  Dimension 5 operators \cite{operators} are of the form
$ \frac{c^2}{M^{eff}_T} \left( ( Q \; Q\; Q\; L ) + (U^c \; U^c\;
D^c\; E^c) \right)$ or commonly described as  L L L L + R R R R
operators.

$\bullet$   Dimension 4 operators given by $(U^c \; D^c \; D^c)  +
(Q \; L \; D^c)  +  (E^c \; L \; L)$ are very dangerous
\cite{operators}.  Fortunately the symmetry $R_p$ or R parity
forbids all dimension 3 and 4 (and even one dimension 5) baryon
and lepton number violating operators.  It is thus a necessary
ingredient of any ``natural" SUSY GUT.

In summary, dimension 4, 5 and 6 operators may contribute to
nucleon decay.  The proton lifetime as a result of dimension 6
operators is very long due to the large value of $M_G$
\cite{susygut}. Dimension 4 operators are necessarily forbidden by
incorporating R parity.  We are thus lead to consider dimension 5
operators which are the dominant contribution to nucleon decay in
SUSY GUTs \cite{dimfiveops}.

The proton decay amplitude depends on four main theoretical
factors.  We have
\begin{eqnarray} T(p \rightarrow K^+ + \bar \nu) \propto &
\frac{c^2}{M^{eff}_T} {\rm (Loop \; Factor) \; (RG)} \; \langle
K^+  \bar \nu | q q q l | p \rangle & \\ & \sim
\frac{c^2}{M^{eff}_T} {\rm (Loop \; Factor) \; (RG)} \;
\frac{\beta_{lattice}}{f_K} \ m_p &\nn \end{eqnarray} where the
latter equation results from a chiral Lagrangian analysis.  Let's
consider each factor in detail\cite{dmr}.

\subsection{$\beta_{lattice}$}   $\beta_{lattice}$ is a 3 quark, strong interaction
matrix element between the vacuum and a nucleon.  A recent lattice
calculation gives\cite{beta}
\begin{equation} \beta_{lattice} = \langle 0 | q q q | N \rangle =
 0.015 (1) \; {\rm GeV}^3 . \end{equation}
 Without going into detail there is also an $\alpha_{lattice}$
 satisfying $\alpha_{lattice} \approx - \beta_{lattice}$\cite{beta}.
 NOTE, in previous theoretical analyses, a conservative lower bound
 $\beta_{lattice} > 0.003 {\rm GeV}^3$ has been used to obtain an
 upper bound on the proton lifetime.   The new lattice result is
 $5 \times$  larger than this  ``conservative lower bound."  This
 has the effect of decreasing the upper bound on the proton
 lifetime by a factor of 25.   The error on the lattice result represents
 statistical errors only.  Systematic uncertainties due to quenching and
 using a chiral Lagrangian analysis may be as large as $\pm 50$ \% (my
 estimate).

\subsection{$c^2$} $c^2$ is a model dependent factor; calculable within a theory of
fermion masses and mixing angles. Dimension 5 baryon and lepton
number violating operators due to color triplet Higgsino exchange
are derived from the superpotential \begin{eqnarray} W \supset &
H_u \; Q Y_u \overline U + & H_d ( Q Y_d \overline D + L Y_e
\overline E) \\ & + T ( Q {1\over 2} c_{qq} Q + \overline U c_{ue}
\overline E ) + &  \bar T ( Q c_{ql} L + \overline U c_{ud}
\overline D ) \nonumber
\end{eqnarray}
obtained by integrating out the Higgs color triplets $( T, \ \bar
T )$ as in the figures below.
\begin{center}
\SetPFont{Helvetica}{15}
\begin{picture}(300,200)(0,0)
\SetColor{Red} \ArrowLine(85,150)(135,150)
\ArrowLine(180,150)(135,150) \PText(115,155)(0)[cb]{T}
\PText(150,155)(0)[cb]{T} \Line(148,169)(154,169)

\SetColor{Green} \ArrowLine(85,150)(30,180)
\ArrowLine(85,150)(30,120) \ArrowLine(180,150)(230,180)
\PText(20,180)(0)[rc]{Q} \PText(20,120)(0)[rc]{Q}
\PText(240,180)(0)[lc]{Q}

\SetColor{Blue} \ArrowLine(180,150)(230,120)
\PText(240,120)(0)[lc]{L}


\SetColor{Green} \ArrowLine(150,50)(90,80)
\ArrowLine(150,50)(90,20) \ArrowLine(150,50)(210,80)
\PText(80,80)(0)[rc]{Q} \PText(80,20)(0)[rc]{Q}
\PText(220,80)(0)[lc]{Q}

\SetColor{Blue} \ArrowLine(150,50)(210,20)
\PText(220,20)(0)[lc]{L}

\SetColor{Red} \Vertex(150,50){3} \Line(20,50)(50,50)
\PText(35,52)(0)[cb]{1} \PText(35,47)(0)[ct]{M}
\PText(48,38)(0)[ct]{T} \PText(50,53)(0)[ct]{eff}

\SetColor{Black}
\end{picture}
\end{center}
We then obtain
\begin{equation} W \supset   H_u \; Q \ Y_u \ \overline U + H_d \
( \ Q \ Y_d \
 \overline D + L \ Y_e \ \overline E) \ +
{1\over{\mt}^{\phantom{(}}} \; \left[ \ Q \ {1\over 2} c_{qq} \ Q
\, Q \ c_{ql} \ L + \overline U \ c_{ud} \ \overline D \ \overline
U \ c_{ue} \ \overline E \ \right]. \end{equation}

The matrix structure of the factor $c^2$ depends on a theory of
charged fermion masses\cite{lucas1,brt,bpw,afm,ab}. For example,
in any realistic GUT model which fits charged fermion masses we
have either, for $SU_5$, an effective superpotential term $W
\supset \lambda(\langle \Phi \rangle) \; 10 \; 10 \; 5_H +
\lambda^\prime(\langle \Phi \rangle) \; 10 \; \bar 5 \; \bar 5_H $
or, for $SO_{10}$, $W \supset \lambda(\langle \Phi \rangle) \; 16
\; 16 \; 10_H $, where $\langle \Phi \rangle$ represents the
vacuum expectation value [vev] of scalars in non-trivial GUT
representations. These vevs are absolutely necessary in order to
fix bad GUT mass relations as discussed below.   As a consequence,
the 3 $\times$ 3 Yukawa matrices $\{ Y_u, \ Y_d, \ Y_e \}$ and the
$c$ matrices $\{ c_{qq}, \ c_{ql}, \ c_{ud}, \ c_{ue} \}$ are
related by GUT symmetry relations.  But, in general, $Y_u \neq
c_{qq} \neq c_{ue}$ and $Y_d \neq Y_e \neq c_{ud} \neq c_{ql}$.

As noted above, the effective superpotential with scalar vevs
$\langle \Phi \rangle$ are needed to correct bad GUT Yukawa
relations.   For example, consider the good GUT relation
$\lambda_b = \lambda_\tau$.  Assuming it also works for the first
two families gives $\lambda_s = \lambda_\mu$ and $\lambda_d =
\lambda_e$. Combining the two we find $20 \sim \frac{m_s}{m_d} =
\frac{m_\mu}{m_e} \sim 200$ which is a bad mass relation.

In general we also have
 \begin{itemize} \item $ \{ c_{qq} \
c_{ql}, \; c_{ud} \ c_{ue} \} \propto m_u \ m_d \ \tan\beta$.

\item If, for example in $SO_{10}$, we allow for Higgs in $16_H$
and $10_H$ which mix and in addition we have higher dimension
operators such as $\frac{1}{M} (16 \ 16 \ 16_H \ 16_H)$ for
neutrino masses then this can lead to new proton decay
operators\cite{bpw}. However this is not required for neutrino
masses (see \cite{brt}).

\item Finally, family symmetries affect the texture of $ c_{qq}, \
c_{ql}, \ c_{ud}, \ c_{ue} $.
\end{itemize}

\subsection{Loop Factor} The Loop Factor depends on the squark, slepton and gaugino
spectrum.  For large $\tan\beta$, it has been shown that proton
decay is significantly constrained by RRRR
operators\cite{lucas1,goto,bs}. The dominant decay mode for the
proton, $p \rightarrow K^+ + \bar \nu_\tau$, is given by the
following graph.
\begin{center} \SetPFont{Helvetica}{15}

\begin{picture}(300,200)(0,0)

\SetColor{Magenta} \ArrowLine(40,160)(100,160)
\PText(30,160)(0)[rc]{s}

\SetColor{Black} \ArrowLine(40,40)(100,40) \Text(30,40)[1c]{$\nu$}
\Text(37,33)[1c]{$\tau$}

\SetColor{Blue} \ArrowLine(220,160)(160,100)
\ArrowLine(220,40)(160,100) \PText(230,40)(0)[lc]{d}
\PText(234,164)(0)[rc]{u}

\SetColor{Brown} \DashArrowLine(100,160)(160,100){5}
\DashArrowLine(100,40)(160,100){5} \PText(130,150)(0)[lc]{t}
\Photon(130,156)(138,158){2}{1} \Text(130,50)[lc]{$\tau$}
\Photon(130,57)(138,59){2}{1}

\SetColor{Green} \Line(100,40)(100,160)
\Photon(100,40)(100,160){7}{4.5} \PText(80,100)(0)[rc]{H}
\Photon(72,105)(80,107){2}{1}

\SetColor{Magenta} \ArrowLine(130,180)(40,180)
\PText(35,190)(0)[rc]{u}

\SetColor{Blue} \ArrowLine(220,180)(130,180)
\PText(225,190)(0)[lc]{u}

\SetColor{Red} \Vertex(160,100){3}

\SetColor{Black}
\end{picture}
\end{center}
This leads to a Loop Factor given approximately by
$\frac{\lambda_t \; \lambda_\tau}{16 \pi^2} \frac{\sqrt{\mu^2 +
M_{1/2}^2}}{m_{16}^2}$.  Minimizing the Loop Factor requires
taking the limit $\mu, M_{1/2} \;\; $ SMALL; $\;\;\; m_{16} \;\; $
Large.

Is this limit reasonable and what about  ``Naturalness"
constraints??  In partial answer to this question let's consider
two additional motivations for being in this particular region of
SUSY breaking parameter space.

An inverted scalar mass hierarchy [with heavy 1st \& 2nd
generation squarks and sleptons ($>>$ TeV) and light 3rd
generation scalars ($\leq$ TeV)] is useful for ameliorating the
SUSY CP and flavor problems.   Such a hierarchy can be obtained
``naturally" via renormalization group running from $M_G$ to $M_Z$
if one assumes the following boundary conditions at the GUT
scale\cite{baggeretal}. One needs a universal A parameter $A_0$,
gaugino mass $M_{1/2}$, Higgs mass $m_{10}$ and squark and slepton
masses $m_{16}$, consistent with $SO_{10}$ boundary conditions. In
addition they must satisfy the constraint: $A_0^2 = 2 m_{10}^2 = 4
m_{16}^2,\;\;\; m_{16} >> 1$ TeV.

On the other hand we can assume $SO_{10}$ Yukawa unification with
$\lambda_t = \lambda_b = \lambda_\tau = \lambda_{\nu_\tau} =
\lambda$ at $M_G$ and see if consistency with the low energy data
constrains the SUSY breaking parameter space\cite{bdr}. Good fits
require the weak scale threshold correction to the bottom quark
mass satisfy $\delta m_b/ m_b = \Delta m_b^{\tilde g} + \Delta
m_b^{\tilde \chi} + \Delta m_b^{\log} + \cdots <  - 2\%$ with
$\Delta m_b^{\tilde g} \approx \frac{2 \alpha_3}{3 \pi} \;
\frac{\mu m_{\tilde g}}{m_{\tilde b}^2} \; tan\beta $ and $ \Delta
m_b^{\tilde \chi^+} \approx \frac{\lambda_t^2}{16 \pi^2} \;
\frac{\mu A_t}{m_{\tilde t}^2} \; tan\beta $.  Requiring $\mu >
0$, which is preferred by $b \rightarrow s \gamma$ and
$a_\mu^{NEW}$, we find that Yukawa unification is possible only in
a narrow region of SUSY parameter space given by $A_0 \sim - 1.9
\; m_{16}$, $m_{10} \sim 1.35 \; m_{16}$, $m_{16} > 1200 \; {\rm
GeV}$ and $\mu, M_{1/2} \sim 100 - 500$ GeV.  Moreover the fits
improve as $m_{16}$ increases.   So from a completely independent
perspective we find the same region of SUSY breaking parameter
space.  Perhaps there is something to this?

In this region of parameter space we find the
predictions\cite{bdr}
\begin{itemize}
\item $\tan\beta \sim 50$; \item $m_{\tilde t} << m_{\tilde b}$;
\item $m_{h} \sim 114 \pm 5 \pm 3$ GeV, and \item $a_\mu^{SUSY} <
16 \times 10^{-10}$
\end{itemize}

\subsection{$M^{eff}_T$} $M^{eff}_T$ is an effective color triplet Higgs mass
which is intimately connected to doublet-triplet splitting and GUT
symmetry breaking.  It may be constrained by requiring
perturbative corrections to the prediction for gauge coupling
unification\cite{lucas2,goto,dmr,bpw,mp,afm}.

Recall the GUT threshold correction $\epsilon_3 \equiv
\frac{(\alpha_3(M_G) - \tilde \alpha_G)}{\tilde \alpha_G} \sim -
4\%$ needed to fit the low energy data. This correction has two
main contributions given by $\epsilon_3 = \epsilon_3^{\rm Higgs} +
\epsilon_3^{\rm GUT \; breaking} + \cdots$.   The Higgs
contribution, in minimal models, is of the form $\epsilon_3^{\rm
Higgs} = \frac{3 \alpha_G}{5 \pi} \ln{\frac{M^{eff}_T}{M_G}}$.  In
the following table, we list the values obtained from the GUT
symmetry breaking sectors of the theory.   Note, minimal $SU_5$
has a negligible contribution from the GUT breaking sector.  Hence
in order to fit the low energy data, a very low value of
$M^{eff}_T$ is needed.  As a result, minimal $SU_5$ is excluded by
proton decay\cite{goto,mp}.  In the following figure, taken from
the paper of Goto and Nihei\cite{goto}, it is clear that minimal
$SU_5$ is excluded by Super-Kamiokande bounds on the proton
lifetime.  Murayama and Pierce\cite{mp} have shown that this
result cannot be saved by an inverted scalar mass hierarchy.
\begin{figure}
\begin{center}
\centerline{ \psfig{file=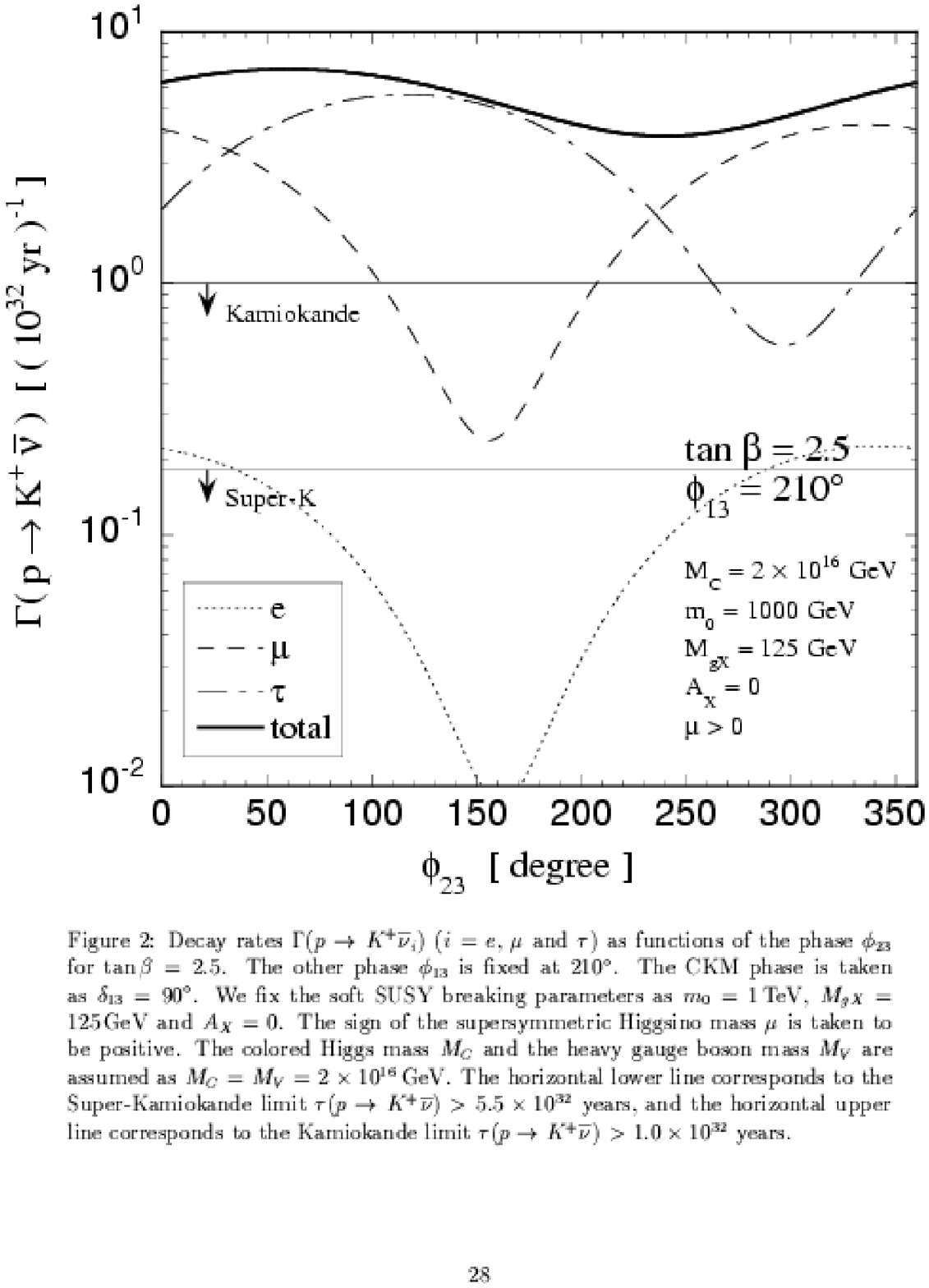,
width=15cm}}
\end{center}
\end{figure}
In the case of $SU_5$ with ``natural" doublet-triplet
splitting\cite{afm} and minimal $SO_{10}$\cite{lucas2}, on the
other hand, it is possible to get significant corrections from the
GUT breaking sector. For $SO_{10}$ we have taken a 10 \%
correction as an (albeit ad hoc) upper limit consistent with
perturbativity\cite{dmr}.

$$\begin{array}{|l|c|c|c|}
\hline
{\rm Model}  &   {\rm Minimal}   &  SU_5  & {\rm Minimal}  \\
&  SU_5  & {\rm ``Natural" D/T} & SO_{10} \\
\hline
\epsilon_3^{\rm GUT breaking} & \approx 0  &  - 7.7 \%  &  -10 \% \\
\epsilon_3^{\rm Higgs} &  - 4 \% & + 3.7 \% &  + 6 \% \\
\hline M^{eff}_T {\rm [GeV]} &  2 \times 10^{14} & 3 \times
10^{18} & 6 \times
10^{19} \\
\hline
\end{array}$$

Note, that in some cases the value of $M^{eff}_T$ is very large.
However in these cases, there is no particle with mass greater
than $M_G$.  By definition we have $\frac{1}{M^{eff}_T} = (
M_T^{-1} )_{11}$ where $M_T$ is the Higgs color triplet mass
matrix.  In the cases of natural doublet-triplet splitting in $SU_5$ and
minimal $SO_{10}$ we have $M_T$ schematically given by
$M_T = \left( \begin{array}{cc} 0 &  M_G \\
M_G & X \end{array} \right)$ with $\frac{1}{M^{eff}_T} \equiv
\frac{X}{M_G^2}$.  Hence for $X << M_G$ we obtain $M^{eff}_T
>> M_G$ and the heaviest Higgs has mass of order $M_G$.

To summarize,  given a realistic GUT with a small set of effective
parameters at the GUT scale, we use the following procedure to
evaluate the proton lifetime\cite{dmr}.  We first vary the
parameters at the GUT scale -- $\tilde \alpha_G, \; M_G$, the $3
\times 3$ Yukawa matrices $Y_u, \; Y_d, \; Y_e$ and the soft SUSY
breaking parameters until we FIT the precision electroweak data,
{\it including fermion masses and mixing angles}.   As a result of
this analysis, $ c_{qq}, \ c_{ql}, \ c_{ud}, \ c_{ue} $ at $M_G$
are FIXED. We then renormalize the dimension 5 operators from $M_G
\rightarrow M_Z$ within the MSSM; evaluate the Loop Factor at
$M_Z$ and renormalize the effective dimension 6 operator from $M_Z
\rightarrow 1$ GeV. The latter gives the renormalization constant
$A_3 \sim 1.3$ (NOT $A_L \sim .22$)\cite{dmr}. Finally we
calculate the decay amplitudes using a chiral Lagrangian approach
or a direct lattice gauge calculation of the appropriate matrix
elements.  As discussed above:
\begin{itemize}
\item $c^2$ is model dependent but constrained by fermion masses
and mixing angles.

\item $\beta_{lattice}$ given by the recent JLQCD central value is
5 times larger than previous ``conservative lower bound".  The
systematic uncertainties of quenching and chiral Lagrangian
analyses need to be evaluated.

\item Loop Factor $\propto \frac{\alpha}{4 \pi} \frac{\sqrt{\mu^2
+ M_{1/2}^2}}{m_{16}^2}$ $\Longrightarrow$  Gauginos light; 1st \&
2nd generation squarks and sleptons $>$ TeV and ``naturalness"
requires the stops, sbottoms and stau mass $<$ 1 TeV.

\item  Finally, $M^{eff}_T$ is constrained by gauge coupling
unification and the GUT breaking sectors.
\end{itemize}

\subsection{The Bottom Line}

\begin{itemize}
\item The proton lifetime due to dimension 6 operators is in the
range $\tau(p \rightarrow e^+ + \pi^0) \sim  10^{35 \pm 1}$ yrs.

 \item The ``upper bound" from dimension 5 operators is roughly
 given by $\tau(p \rightarrow K^+ + \bar \nu) < (\frac{1}{3} - 3) \times
10^{34}$ yrs\cite{dmr,bpw,afm}.

\item  In general, $\tau(n \rightarrow K^0 + \bar \nu) < \tau(p
\rightarrow K^+ + \bar \nu)$, and

\item Other decay modes may be significant, eg. $p \rightarrow
K^0 +  \mu^+$,  $ \pi^0 +  e^+$, but this is very model dependent.
\end{itemize}

Can we eliminate dimension 5 operators by symmetries?   This is
non-trivial in 4 dimensions, but possible\cite{babubarr}.
 It is however ``natural" in extra dimensions with GUT symmetry breaking by
orbifold boundary conditions\cite{extradim}.

\section{5 Dim. SUSY GUTs} Let's consider one recent construction
of a complete $SU_5$ SUSY GUT in five dimensions\cite{5dim}.  The
picture below represents flat 3 + 1 dimensional end of the world
branes separated by a fifth dimensional line segment which runs
from 0 to $\pi R$.  The fifth dimension is an orbifold which has
an $SU_5$ gauge symmetry on the 0 Brane and in the bulk. However
the symmetry on the $\pi R$ Brane is only the standard model $SU_3
\times SU_2 \times U_1$.  Quarks and leptons in the third family
sit on the 0 Brane as indicated in the figure. On the other hand,
the quarks and leptons in the first two families are partially on
the 0 Brane and partially in the bulk. Finally the two Higgs
doublets sit in the bulk.
\begin{center}
\SetPFont{Helvetica}{15}
\begin{picture}(300,100)(0,0)
\SetColor{Red} \Line(95,90)(95,20) \Line(70,100)(70,15)
\Line(95,90)(70,100) \Line(95,20)(70,15)

\SetColor{Black} \DashLine(70,15)(160,15){2} \Text(70,5)[1c]{$0$}
\Text(160,5)[1c]{$\pi$R} \Text(40,60)[1c]{$SU_5$}
\Text(220,60)[1c]{$SU_3 \times SU_2 \times U_1$}

\SetColor{Blue} \Text(130,90)[1c]{Gauge} \Text(130,70)[1c]{$5_H +
5_H^c$} \Text(130,50)[1c]{$\bar 5_H + \bar 5_H^c$}
\Text(130,30)[1c]{$10_{1,2}$} \Text(80,70)[1c]{$10_3$}
\Text(82,50)[1c]{$\bar 5_{1,2,3}$}

\SetColor{Green} \Line(185,90)(185,20) \Line(160,100)(160,15)
\Line(185,90)(160,100) \Line(185,20)(160,15)

\SetColor{Black}
\end{picture}
\end{center}
In order to preserve the predictions of gauge coupling unification
whereby bulk symmetry relations dominate over the broken symmetry
on the $\pi R$ Brane one must have a ``large" extra dimension with
a compactification scale $M_C$ satisfying $M_C = \frac{1}{R} \sim
10^{15}$ GeV $ << M^*$ with $M^* \sim 10^{17}$ GeV, the cut-off
scale. The three gauge couplings unify at $M^*$, but the baryon
and lepton number violating gauge bosons obtain mass at $M_C$.

The resulting 5d theory has the following virtues:
\begin{itemize}
\item Threshold corrections to gauge coupling unification, coming
from Kaluza-Klein modes with mass between $M_C$ and $M^*$, are
just right to give a good fit to the low energy data.

\item The theory preserves the good Yukawa relation: $\lambda_b =
\lambda_\tau$.

\item  The orbifold symmetry breaking results in ``natural"
doublet-triplet Higgs splitting with a conserved $U_1(R)$
symmetry.

\item R parity ($R_p$) is a subgroup of $U_1(R)$ and is thus
conserved.

\item Dimension 5 baryon number violating operators are
eliminated.
\end{itemize}

On the other hand, \begin{itemize}

\item there is NO explanation of charge quantization, since weak
hypercharge is not quantized on the $\pi R$ Brane.

\item Since the physical quarks and leptons, coming from bulk
states, are derived from many different $10$s the GUT explanation
for families of quarks and leptons is lost.

\item Proton decay due to dimension 6 operators may be enhanced.
The decay amplitude depends on new effective operators with an
unknown dimensionful coupling given by $\frac{\bar g}{M_C^2}$.
Assuming $\bar g \sim 1$, we obtain a proton lifetime  $\tau(p
\rightarrow \pi^0 + e^+) \sim 10^{34}$ yrs.
\end{itemize}

\section{Conclusions}

4D SUSY GUTs are still very much alive.
\begin{itemize}
\item The proton lifetime due to dimension 6 operators is in the
range $\tau(p \rightarrow e^+ + \pi^0) \sim  10^{35 \pm 1}$ yrs.

\item The ``upper bound" due to dimension 5 operators is given by
$\tau(p \rightarrow K^+ + \bar \nu) < (\frac{1}{3} - 3) \times
10^{34}$ yrs.

\item  Suppressing proton decay due to dimension 5 operators
requires light gauginos and heavy 1st \& 2nd generation squarks
and sleptons.

\subitem This range of soft SUSY breaking parameters is consistent
with $SO_{10}$ Yukawa unification which also favors $m_{h} \sim
114 \pm 5 \pm 3$ GeV.

\subitem It is also consistent with an inverted scalar mass
hierarchy which ameliorates the SUSY CP and flavor problem.

\subitem Finally for $m_{16} > 1200$ GeV, we find $a_\mu^{SUSY} <
16 \times 10^{-10}$.
\end{itemize}

\bigskip

SUSY GUTs in Extra Dimensions

\begin{itemize}
\item There are NO miracles in extra dimensions.

\item Dimension 5 operators can easily be eliminated in extra
dimensions {\it at the expense of NOT understanding charge
quantization.}

\item In 5D GUTs,  we can have ``natural" doublet-triplet Higgs
splitting and a conserved $R_p$.

\item Finally, proton decay due to dimension 6 operators may be
enhanced giving $\tau(p \rightarrow \pi^0 + e^+)$ roughly of order
$10^{34}$ yrs.
\end{itemize}

In conclusion, SUSY GUTs in four or even higher dimensions lead to
proton decay rates which may easily be observable in a future
proton decay experiment.   Moreover an observation of proton
and/or neutron decay would provide a tantalizing window to new
physics ``way" beyond the standard model.

%

\end{document}